\documentclass[final,5p,times,twocolumn]{elsarticle}
\usepackage{graphicx}
\usepackage{slashed,color}
\usepackage{amsmath,amsfonts,bm}
\usepackage{amssymb}
\usepackage{calligra}
\usepackage[T1]{fontenc}

\newcommand{\nn}{\nonumber \\ }

\journal{Physics Letters  B}

\begin{document}



\author{A.~V.~Radyushkin}
\address{Physics Department, Old Dominion University, Norfolk,
             VA 23529, USA}
\address{Thomas Jefferson National Accelerator Facility,
              Newport News, VA 23606, USA
}

\title{Nonperturbative Evolution of Parton Quasi-Distributions }

\begin{abstract}

Using  the   formalism of parton virtuality distribution  functions (VDFs) 
we establish a connection between the 
transverse momentum dependent 
 distributions (TMDs) ${\cal F} (x, k_\perp^2)$ and 
quasi-distributions  (PQDs) $Q(y,p_3)$
introduced recently  by X. Ji for lattice QCD extraction of parton distributions $f(x)$.
We build  models for PQDs from  the VDF-based models for soft TMDs,  and  analyze 
the $p_3$ dependence of  the resulting PQDs. We observe  a strong 
nonperturbative evolution of PQDs for small and moderately large values of $p_3$
reflecting the transverse momentum dependence of TMDs. 
Thus,  the study of  PQDs on the lattice in the domain 
of strong nonperturbative effects  opens a new perspective for  investigation of  the 
3-dimensional hadron structure.  


              
\end{abstract}

\maketitle


   
\section{Introduction}

The parton distribution functions (PDFs) $f(x)$, being related to   matrix elements
of nonlocal operators near  the light cone $z^2=0$ are notoriously 
difficult objects for a calculation using  the  lattice gauge  theory.
The latter is formulated in the Euclidean space where  light-like separations do not exist.
Recently, X. Ji  \cite{Ji:2013dva} proposed to use purely space-like
separations $z=(0,0,0,z_3)$ to   overcome this problem.

The parton {\it quasi-distributions}   $Q(y,p_3)$ introduced by X. Ji,
differ from  PDFs $f(x)$, but tend to them in the \mbox{$p_3 \to \infty$}  limit,
displaying a usual perturbative evolution 
\mbox{\cite{Gribov:1972ri} -- \cite{Dokshitzer:1977sg} } 
with respect to $p_3$  for large $p_3$. 
Refs. \cite{Ji:2013dva}, \cite{Xiong:2013bka} -- \cite{Ji:2015qla}
discuss the properties of PQDs in the large $p_3$ limit and their  matching  
with  scale-dependent PDFs $f(x,\mu)$.  
The results of lattice calculations of PQDs were reported  
in Refs. 
\cite{Lin:2014yra} -- \cite{Alexandrou:2016tjj}.

These results  show a significant variation of PQDs with $p_3$.
However, since the values of $p_3$ used in these calculations are not very
large,  the observed $p_3$   evolution does not have a perturbative form.
The nonperturbative aspects of the $p_3$-evolution were studied 
in diquark spectator  models \cite{Gamberg:2014zwa,Gamberg:2015opc,Bacchetta:2016zjm} 
for parton distributions. The evolution patterns observed in these papers
are in a qualitative agreement with the lattice results. 
The authors also discuss the $p_3 \to \infty$ extrapolation of results obtained 
for moderately large $p_3$  values. 

Our goal in the present paper  is to study nonperturbative evolution of parton
quasi-distributions using the formalism of {\it virtuality distribution functions} 
  proposed   and developed 
   in our recent papers  \cite{Radyushkin:2014vla,Radyushkin:2015gpa},
   where it was applied to the transverse momentum dependent pion 
   distribution amplitude and 
  the exclusive $\gamma^* \gamma \to \pi^0$ 
 process.   
 
 To this end,   in Section  2  we   extend the VDF formalism onto 
 the parton distribution functions, and show how 
 the  basic  VDF $\Phi (x, \sigma)$ is related to PDFs,
 to TMDs 
 and to PQDs.    In particular, we show that PQDs are completely determined
 by TMDs through a rather simple transformation.
 Since the basic relations between the parton distributions
 are rather insensitive to complications brought by spin,
 in Section  2 we refer to a simple scalar model.
 In Section  3 we discuss modifications related to  quark spin 
 and gauge nature of gluons in quantum chromodynamics (QCD).
 In Section 4 we discuss VDF-based models for soft TMDs,
 and in Section 5 we present our results for  nonperturbative evolution 
 of PQDs obtained in these models. The transition to perturbative evolution for large $p_3$
 is discussed in Section  6. 
 Our conclusions are given in Section 7.

 \setcounter{equation}{0}   \section{Parton distributions} 


\subsection{Virtuality distribution functions}

  Historically, parton distributions  \cite{Feynman:1973xc} were  introduced
  to describe inclusive deep inelastic scattering   involving spin-1/2 quarks. 
Since  complications related to spin
 do not affect  
  the very concept of parton distributions,  we  start with 
  a simple example of a  scalar  theory.
Then information about the target is accumulated 
in the generic matrix element $\langle  p | \phi (0)  \phi(z)  | p \rangle $.
Transforming to
 the   momentum space 
\begin{align}
 \langle  p | \phi (0)  \phi(z)  | p \rangle = \frac{1}{\pi^2}  \int { d^4 k}  \, e^{- ikz} \,   \chi (k,p)  
 \label{twist2parz}
\end{align}
we switch to  the description in terms of   $\chi (k,p)$  which  is an analog of 
the Bethe-Salpeter  amplitude \cite{Salpeter:1951sz}.

      \begin{figure}[t]
    \centerline{\includegraphics[width=2.5in]{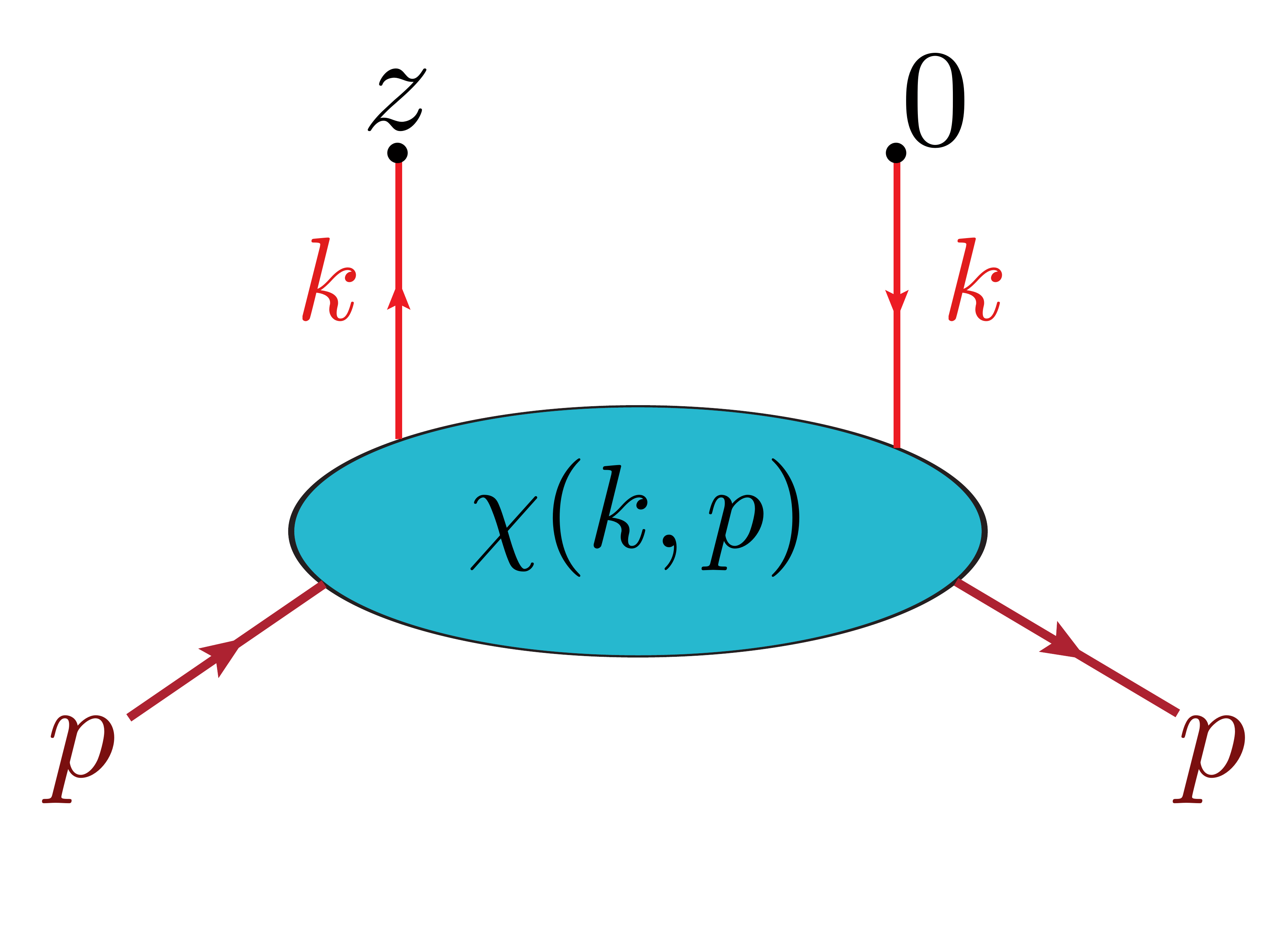}}
    \caption{Structure of parton-hadron matrix element.
    \label{chipk}}
    \end{figure}

 A crucial observation is that  the contribution of 
 any  (uncut) diagram  to $\chi (k,p) $
 may be written as 
\begin{align} 
& i \chi (k,p)  = \int_{0}^{\infty} d\lambda  \,  \int_{-1}^{1} d x \,
e^{i  \lambda[  k^2 - 2x (kp) +i\epsilon]} F(x, \lambda;M^2)  
\label{chisca} 
\end{align}

 The  reason is that 
for a general scalar 
handbag diagram  $d_i$ one can write (see, e.g., \cite{Nakanishi:1971graph}) 
\begin{align}
  &i  \chi_{d_i}  (k,p) =  i^{l} \, \frac{P({\rm c.c.})}{(4\pi i)^{2L}}
\int_0^{\infty} \prod_{j=1}^l   d\alpha_{j} [D(\alpha)]^{-2}
\nonumber \\ & \times 
\exp \left \{ i k^2  \frac{A (\alpha)}{D(\alpha) } +i 
  \frac{ (p-k)^2 B_s (\alpha)+   (p+k)^2 B_u (\alpha) }{D(\alpha) }
 \right \} 
\nonumber \\ & \times 
\exp \left \{ 
i  M^2  \frac{ C(\alpha) }{D (\alpha) }
- i  \sum_{j} \alpha_{j} (m_{j}^2- i\epsilon) \right \}  \ , 
\label{alphap}
\end{align}
where  $M^2=p^2$, ${P({\rm c.c.})}$ is the relevant
 product of  coupling constants, 
 $L$ is the number of loops of the diagram,  and $l$
is the number  of its
 lines.  For our purposes,
the most important property of this representation is that 
$A(\alpha), B_s(\alpha),  B_u(\alpha),  C(\alpha), D(\alpha)$ 
are positive (or better, non-negative) functions (sums of products) 
of the non-negative $\alpha_j$-parameters 
of a  diagram. 
Using it, we get the representation 
(\ref{chisca})  with 
\begin{align}
\lambda= & \frac{A(\alpha)+ B_s(\alpha) +  B_u(\alpha)}{D(\alpha) } \  \  ,  \label{lambda}  \\ 
 x= & \frac{B_s(\alpha) -  B_u(\alpha)}{A(\alpha)+ B_s(\alpha) +  B_u(\alpha)} \ ,
 \label{xal}
  \end{align}
and
a  function $F(x, \lambda;M^2)$  specific 
for each diagram. 
Evidently, 
$0\leq \lambda \leq\infty$. 
The limits for $x$ in  general case are \mbox{$-1\leq x\leq 1$, } 
the negative $x$ appearing when $B_u (\alpha) \neq 0$,  
which happens for some   nonplanar diagrams. 

    Integrating over $\lambda$ in Eq.~(\ref{chisca})  gives  
      a Nakanishi-type  representation (see, e.g. \cite{Nakanishi:1969ph}) for this amplitude.
We prefer, however, to  use  the representation 
involving both $x$ and $\lambda$ as integration variables. 

Note that {\it no} restrictions (like being lightlike, etc.)
are imposed on $k$ and $p$ in Eq. (\ref{chisca}).
In particular, $p$ is the actual external momentum with $p^2=M^2$. 
Basically, Eq. (\ref{chisca}) expresses an  obvious  fact that, 
due to the Lorentz invariance, the function $\chi (k,p)$ 
depends on $k$ through 
 $(kp)$ and $k^2$. It may be treated as      a double Fourier representation  
of $\chi (k,p)$ in 
both $(kp)$ and $k^2$.

 Transforming  Eq.  (\ref{chisca})   to the coordinate 
representation and changing $\lambda =1/\sigma$  gives
 \begin{align}
  \langle p |   \phi(0) \phi (z)|p \rangle 
=  & \int_{0}^{\infty} {d\sigma} \,
  \,   \int_{-1}^{1} d x \, 
  e^{-i x (pz) -i \sigma {(z^2-i \epsilon )}/{4}} \nn & \times 
e^{-i  x^2 M^2/\sigma} F  (x, 1/\sigma;M^2)
	  \,  . 
 \
 \label{newTpqx2}
\end{align} 
Defining  the  {\it Virtuality Distribution Function} 
 \begin{align}
\Phi  (x, \sigma; M^2) = 
\exp [-ix^2  M^2/\sigma] F  (x, 1/\sigma; M^2)
\label{Phidef}
\end{align} 
we arrive at the  {\it VDF representation}  
 \begin{align}
 \nonumber 
  \langle p |   \phi(0) \phi (z)|p \rangle 
=  & 
\int_{0}^{\infty} d \sigma \int_{-1}^1 dx\,  %
 \Phi (x,\sigma; M^2) \, \\ & \times \,  e^{-i x (pz) -i \sigma {(z^2-i \epsilon )}/{4}} \,  
 \
 \label{newVDFx}
\end{align} 
that reflects the fact that the matrix element 
$\langle p |   \phi(0) \phi (z)|p \rangle$  depends on $z$ through 
$(pz)$ and $z^2$, and may be treated as a double Fourier representation 
with respect to these variables.
On general grounds, one would expect that such a
Fourier  representation
should be valid for a very wide class of functions.
The main 
     non-trivial   feature of the representations (\ref{chisca}),   (\ref{newVDFx})
is  in their  specific  limits of integration over  $x$ and $\lambda$ (or $\sigma$).  
For an arbitrary   function, one cannot insist on such  limits.

However, our   matrix element is not an
 arbitrary function. It is given by a sum of 
  handbag Feynman diagrams, and the limits on $x$ and
 $\lambda$ (or $\sigma$) are dictated by the properties of these diagrams,
 in particular, by positivity of the functions
 $A,B,D$  determining $x$ and $\lambda$.
 It should be emphasized that  these functions are determined purely 
 by   denominators of propagators, and are not affected by their
 numerators present in non-scalar theories.

 Thus, the VDF representation (\ref{newVDFx})  is valid  for any diagram and 
 reflects very general  features of quantum field theory.  On these  grounds, we 
 will assume that it holds nonperturbatively.
 An important point is that  Eq. (\ref{newVDFx})   gives a covariant definition of $x$ as 
a variable that is Fourier-conjugate to $(pz)$. There is no need to assume that 
$p^2=0$ or $z^2=0$ to define $x$. 
 The parameter $\sigma$,
being conjugate to $z^2$,  may be interpreted as some measure of parton virtuality,
hence the name of the function. 
In particular, 
 VDF  contains higher-twist contributions describing  transverse momentum effects.

\subsection{Collinear   PDFs and TMDs }

While the VDF representation holds for any $z$ and $p$, 
nothing prevents us from considering some special cases,
like a projection on the light cone $z^2=0$. 
This may be implemented, e.g.,  by choosing  $z$
that  has the minus component only. Then one can 
parameterize the matrix element  in terms of  the twist-2 parton distribution $f(x)$
  \begin{align}
  \langle  p |\phi (0) \phi(z_-)  
| p \rangle =
   \int_{-1}^1 dx \, f(x) \, 
e^{-ixp_+ z_-} \,  \  
 \label{twist2par0}
\end{align}
that depends on the fraction $x$ of the target momentum 
component $p_+$ 
carried by the parton. 
The   relation between VDF $\Phi  (x, \sigma) $ and the collinear  twist-2  PDF $f(x)$
is formally given by 
\begin{align} 
\int_{0}^{\infty}  \Phi (x,\sigma) \, d \sigma  = f  (x)   \ .
\label{Phix0}
\end{align} 

  Of course,  this construction of $f(x)$  works only if the 
  $z^2 \to 0$ limit is finite, e.g. in the  super-renormalizable
  $\varphi^3$ theory.  In   the  renormalizable
  $\varphi^4$  theory, the function $ \Phi (x,\sigma)$ has a $\sim 1/\sigma$ hard 
  part, and the integral (\ref{Phix0}) is logarithmically divergent,
  reflecting the perturbative evolution of parton densities in such a theory.

Treating the target momentum $p$ as  purely longitudinal,
$p= (E, {\bf 0}_\perp, P)$,  one can introduce  the parton's
transverse momentum.    In the light-front variables [we use   the 
convention \mbox{$ (ab)=  a_+ b_- + a_- b_+  - a_\perp  b_\perp$}], we write 
    \mbox{$ p=(p_+ , p_- =  p^2/2p_+, p_\perp=0)$} .
Taking $z$ that has 
 $z_-$ and $z_\perp$ components only, i.e.,
 projecting on the light front $z_+=0$,  
 we define   the {\it transverse momentum dependent distribution} 
 in the  usual way 
  as a Fourier transform
with respect to remaining  coordinates $z_-$ and $z_\perp$: 
\begin{align}
{\cal F} (x,k_\perp^2)= &\frac{ p_+}{(2 \pi)^3}  \int_{-\infty}^\infty dz_-
\int {d^2z_\perp}\, e^{i(k_\perp z_\perp )} \, e^{i x z_- p_+}
\nn & \times 
\langle  p | \phi (0)  \phi(z_-, z_\perp)  | p \rangle |_{ p_\perp=0}  \ \ .
\label{TMDchi}
\end{align}
Because of  the rotational 
invariance in $z_\perp$ plane, this   TMD  depends on $k_\perp^2$  only, the 
fact  already  reflected in  the notation. 
The TMD may be written in terms of VDF as  
\begin{align}
&{\cal  F} (x, k_\perp^2) =  \frac{i }{\pi} 
\int_{0}^{\infty} \frac{d \sigma }{\sigma} \, 
 \Phi (x,\sigma) \,  \,  
 e^{- i (k_\perp^2-i \epsilon )/ \sigma} \  .  
\label{TMDsig} 
\end{align}

Note that having a covariantly  defined VDF $ \Phi (x,\sigma)$,
one can use this representation to analytically continue 
${\cal  F} (x, k_\perp^2) $ into a  region of negative and even complex values of
$k_\perp^2$.

The integrated TMD   
 \begin{align}
f(x, \mu^2) \equiv \pi   \int^{\mu^2}_0 d k_\perp^2   {\cal F} (x, k_\perp^2) 
\label{intTMD}
\end{align}  
may be interpreted as a  scale-dependent parton distribution.
Indeed, when   the $\mu^2 \to \infty $ limit  exists, we 
have
$
f (x, \infty) = f(x)  \ . 
$
One can write $f(x, \mu^2)$ in terms of VDF, 
\begin{align}
&f(x, \mu^2)   = 
\int_{0}^{\infty} {d \sigma }\, \left [ 1 - e^{- i (\mu^2-i \epsilon )/\sigma} \right ]\, 
 \Phi (x,\sigma) \,  \  .  
\label{intTMDsig} 
\end{align} 
Note that $f(x, \mu^2)$ has $\mu^2$-dependence, i.e. {\it evolves}
with $\mu^2$ even if the limit $\mu^2 \to \infty$ is finite, e.g. in a super-renormalizable 
theory.  The evolution equation 
\begin{align}
\mu^2 \frac{d}{d \mu^2} f(x, \mu^2)   = & \pi \mu^2  {\cal F} (x, \mu^2) \  
\label{intTMDevo} 
\end{align} 
follows from the definition (\ref{intTMD}). 
When the TMD ${\cal F} (x, k_\perp^2)$  vanishes faster than 
$1/k_\perp^2$ (such a TMD will be called ``soft''),  the evolution 
essentially stops at large $\mu^2$. 

In a renormalizable theory, it makes sense to represent 
$ \Phi (x,\sigma)$  as a sum of a soft part $ \Phi^{\rm soft}  (x,\sigma)$, 
generating a nonperturbative evolution of $f(x,\mu^2)$, 
and a $\sim1/\sigma$ hard tail.  Namely,  the lowest-order hard-tail  term 
 \begin{align}
   \Phi^{\rm hard}  (x,\sigma) \, = {\Delta (x)}/{\sigma} \,  
 \  , 
 \label{Phihard}
\end{align} 
 with  $\Delta (x)$  given by
 \begin{align}
 \Delta (x) =a  \int_x^1 \frac{dz}{z}  \, P(x/z) \, f^{\rm soft}(z) 
 \
 \label{deltaev}
\end{align} 
(where $P(x/z)$ is the evolution kernel
and  $a$  is the appropriate coupling constant) 
 generates the perturbative  evolution
\begin{align}
\mu^2 \frac{d}{d \mu^2} f^{\rm hard} (x, \mu^2)   = & 
a \int_x^1 \frac{dy}{z} P(x/z) \, f^{\rm soft} (z)\  . 
\label{intTMDevohard} 
\end{align} 

The theory of  perturbative evolution (which includes also the subtleties
of using the running coupling constant,  higher-order corrections, scheme-dependence, 
etc.) is well developed, and it is not of much interest for us in this paper.
Our main subject in what follows is the nonperturbative evolution
generated by the soft part of VDF $\Phi (x, \sigma)$ [or TMD ${\cal F} (x, k_\perp^2)$],
in application to  parton quasi-distributions, introduced recently by X. Ji  \cite{Ji:2013dva}.

\subsection{Quasi-Distributions} 

The basic idea of Ref. \cite{Ji:2013dva} is to consider equal-time bilocal operator 
corresponding to 
$z= (0,0,0,z_3)$ [or, for brevity, \mbox{$z=z_3$}]. Then 
 \begin{align}
  \langle p |   \phi(0) \phi (z_3)|p \rangle 
=  & 
\int_{0}^{\infty} d \sigma \int_{-1}^1 dx\,  %
 \Phi (x,\sigma) \, 
 e^{i x p_3 z_3 +i \sigma z_3^2/{4}} \,  .
 \
 \label{newVDFxz3}
\end{align} 
Using  again the   frame in which $p=(E, 0_\perp, P)$, 
 and 
introducing  quasi-distributions \cite{Ji:2013dva}  through 
 \begin{align}
  \langle p |   \phi(0) \phi (z_3)|p \rangle 
=  & 
\int_{-\infty}^{\infty}   dy \, 
 Q(y, P) \,  e^{i y  P z_3 } \, 
 \
 \label{newVDFxzQ}
\end{align} 
we get a relation between PQDs and VDFs,
 \begin{align}
 Q(y, P)  = & \,\int_{0}^{\infty} d   \sigma \sqrt{\frac{i \, P^2}{\pi \sigma}}
 \int_{-1}^1 dx\,  %
 \Phi (x,\sigma) \, 
  e^{- i (x -y)^2 P^2 / \sigma }
 \ . 
 \label{newVDFzQin2}
\end{align} 
For large $P$,
we have 
 \begin{align}
  \sqrt{\frac{i\, P^2}{\pi \sigma}}
  e^{- i (x -y)^2 P^2 / \sigma } =  \delta (x-y) + \frac{\sigma}{4 P^2} \delta'' (x-y) + \ldots
 \
 \label{Qin3}
\end{align} 
and $Q(y, P\to \infty)$ tends to the integral  (\ref{Phix0})  producing $f(y)$. This observation 
suggests that one may be able to  extract the ``light-cone'' parton distribution $f(y)$
from the  studies of  the purely ``space-like''   function  $Q(y, P)$ for large $P$, which 
can be done on the  lattice \cite{Ji:2013dva}. 

The nonperturbative evolution of $Q^{\rm soft} (y,P)$ with respect to $P$ has the area-preserving property.
Namely, since 
 \begin{align}
 \int_{-\infty}^\infty dy\, 
  e^{- i(x -y)^2 P^2 / \sigma } %
  = \sqrt{\frac{\pi \sigma}{i P^2}}
 \ , 
 \label{Qin71}
\end{align} 
we  formally have 
 \begin{align}
 \int_{-\infty}^\infty dy\,  Q(y, P)  = &
\, \int_{0}^{\infty} d   \sigma 
 \int_{-1}^1 dx\,  %
  \Phi (x,\sigma)
   \  .
 \
 \label{Qin810}
\end{align} 
For the soft part,   the integral over $\sigma$ converges, and we may write
 \begin{align}
 \int_{-\infty}^\infty dy\,  Q^{\rm soft} (y, P)  =  \int_{-1}^1 dx\,  
f^{\rm soft} (x)  \  ,
 \
 \label{Qin81}
\end{align} 
which   means that $Q^{\rm soft} (y,P)$  for any $P$ 
has  the same area normalization as $f^{\rm soft}(x)$.  
Note also that the result of Eq. (\ref{Qin81}) 
may be obtained by formally  taking $z_3=0$ in  the definition (\ref{newVDFxzQ})  of $Q(y,P)$.

Similarly, 
since 
 \begin{align}
 \int_{-\infty}^\infty dy\, y\, 
  e^{- i(x -y)^2 P^2 / \sigma } %
  = x \, \sqrt{\frac{\pi \sigma}{i P^2}}
 \ , 
 \label{Qin712}
\end{align} 
we  formally have 
 \begin{align}
 \int_{-\infty}^\infty dy\, y\,  Q(y, P)  = &
\, \int_{0}^{\infty} d   \sigma 
 \int_{-1}^1 dx\,  %
 x \,  \Phi (x,\sigma)
   \  .
 \
 \label{Qin810m}
\end{align} 
  Again, the integral over $\sigma$ converges for the soft part,  and 
we have  the momentum sum rule
 \begin{align}
 \int_{-\infty}^\infty dy\,  y\, Q^{\rm soft} (y, P)  = 
  \int_{-1}^1 dx\,  x \, 
f^{\rm soft} (x)  \  .
 \
 \label{Qin812}
\end{align} 

Finally, comparing the VDF representation (\ref{newVDFzQin2})  for $Q(y, P)$ with that for the TMD
${\cal F} (x, k_\perp^2)$ [see (\ref{TMDsig} )]  we conclude that
  \begin{align}
 Q(y, P)  =  & \,\int_{-\infty}^{\infty} d  k_1
   \int_{-1}^1 dx\, P\,  {\cal F} (x, k_1^2+(x -y)^2 P^2 )
  \  .
 \label{QTMD}
\end{align} 
Thus, the quasi-distribution $ Q(y, P)$ [both its soft and hard parts] 
is  completely determined by the form
of the TMD  ${\cal F} (x, k_\perp^2)$.

 \setcounter{equation}{0} 
\section{QCD}

\subsection{Spinor quarks}

In spinor case, one deals with the matrix element  of a 
    \begin{align}
 B^\alpha  (z,p) \equiv \langle  p | \bar \psi (0)\gamma^\alpha  \psi(z)  | p \rangle \  
\end{align}
type. It may be decomposed into $p^\alpha$ and $z^\alpha$ parts:
$B^\alpha  (z,p) = p^\alpha B_p (z,p) + z^\alpha B_z (z,p)$, or in the VDF 
representation 
\begin{align}
& B^\alpha  (z,p)
  = 
\int_{0}^{\infty} d \sigma \int_{-1}^1 dx\,  \nonumber \\ & \times 
\bigl [ 2 p^\alpha \Phi (x,\sigma) 
+z^\alpha Z (x,\sigma) \bigr ] \, 
 \,  e^{-i  x (pz) -i \sigma {(z^2-i \epsilon )}/{4}} \ . 
 \label{OPhixspin0}
\end{align} 

If we take  $z=(z_-, z_\perp)$ in the $\alpha=+$  component of 
${\cal O}^\alpha$,  the purely higher-twist $z^\alpha$-part drops out 
and we can introduce the TMD  ${\cal F}(x, k_\perp^2)$  that 
is  related to the VDF $\Phi (x,\sigma)$ by the scalar formula 
(\ref{TMDsig}). 

The easiest way to avoid the effects of the $z^\alpha$ contamination in   the quasi-distributions 
 is to take the time component of \mbox{$B^\alpha  (z=z_3,p)$}  and define
 \begin{align}
& B^0   (z_3,p)   
  = 2 p^0 
 \int_{-1}^1 dx\,  
Q(y,P) \, 
 \,  e^{i  y Pz_3 } 
 \label{OPhixspin12}
\end{align} 
(here we differ from the original definition of PQDs by X. Ji \cite{Ji:2013dva}  who uses 
$\alpha=3$). 
The connection between $Q(y,P)$  and $\Phi (x, \sigma)$  is  given then by the 
same formula (\ref{newVDFzQin2}) as in the scalar case.  As a result,
we   have the sum rules  (\ref{Qin81}) and (\ref{Qin812})  corresponding 
to charge and momentum conservation. 
Furthermore, the quasi-distributions $Q_i (y,P)$ are related
to TMDs ${\cal F}_i(x, k_\perp^2)$ by the scalar conversion formula (\ref{QTMD}).

\subsection{Gauge fields}

In QCD, one should take the operator 
\begin{align}
{\cal O}_q^\alpha  (0,z; A) \equiv \bar \psi (0) \,
 \gamma^\alpha \,  { \hat E} (0,z; A) \psi (z)  \  
\end{align}
involving a  straight-line 
 path-ordered exponential
\begin{align}
{ \hat E}(0,z; A) \equiv P \exp{ \left [ ig \,  z_\nu\, \int_0^1dt \,   A^\nu (t z) 
 \right ] }  
 \label{straightE}
\end{align}
in the quark (adjoint) representation. 
As is well-known, its   Taylor expansion   has 
the same structure 
as that for the original $\bar \psi (0)   \gamma^\alpha \psi (z)$ 
operator, with the only change   that
one should use covariant derivatives 
\mbox{$D^\nu =\partial^\nu  - ig A^\nu$}  instead of the
ordinary  $\partial^\nu $ ones:
\begin{align}
  { \hat E} (0,z; A)\,  \psi (z) = \sum_{n=0}^{\infty} 
 \frac{1}{n!} (z D)^n  \psi (0) \ . 
\label{Taylorpsi}
\end{align}

Again, the  $z^\alpha$ contamination is avoided if  the quasi-distributions 
are   defined through the time component of ${\cal O}^\alpha$. 
Then we 
have the same relations between the VDFs and PQDs as in the scalar case.

\subsection{Sum Rules}

Converting  Eq. (\ref{Qin810})  into the sum rule (\ref{Qin81})    
we noted that in general it holds  for the soft part only, 
because the  hard part $\Phi^{\rm hard}  (x,\sigma)$ (\ref{deltaev}) 
is proportional to $1/\sigma$ and its  $\sigma$-integral logarithmically 
diverges.  However,  the $x$-integral of $\Phi^{\rm hard}  (x,\sigma)$ 
 vanishes
(the zeroth $x$-moment  of the evolution kernel $P_{qq} (x/z)$ 
is proportional  to the anomalous dimension of  the vector current, which is zero
due to the   
 vector current conservation).
 As a result, we have the valence  quark  sum rules 
 \begin{align}
 \int_{-\infty}^\infty dy\,  [Q_q(y, P) - Q_{\bar q} (y,P)] =
  \int_{-1}^1 dx\,  
[f_q(x,\mu^2)- f_{\bar q} (x,\mu^2)]  \  
 \
 \label{Qin815}
\end{align} 
involving full PQDs and PDFs. 

Since the first $x$-moment of $P_{qq} (x/z)$ is non-zero, Eq. (\ref{Qin812})  
may be only  used to derive 
the momentum sum rule involving   the 
soft parts of quark distributions  
 \begin{align}
 \int_{-\infty}^\infty dy\,  y\, [Q^{\rm soft} _q(y, P)  + Q^{\rm soft} _{\bar q} (y,P] = 
  \int_{-1}^1 dx\,  x \, 
[f^{\rm soft} _q(x)+ f^{\rm soft} _{\bar q} (x)]  \  . 
 \
 \label{Qin8123}
\end{align}

To include gluons, one should consider the  operator  
\begin{align}
{\cal O}_g^{\alpha \beta}  (0,z; A) \equiv G^{\alpha \nu}  (0) \,
 { \tilde E} (0,z; A) {G_\nu} ^\beta  (z)  \  .
\end{align}
Here $\tilde E$ is   the straight-line 
 path-ordered exponential in the gluon (fundamental) representation.
 The  matrix element  of ${\cal O}_g^{\alpha \beta}  (0,z; A) $  
 contains the basic $p^\alpha p^\beta$ structure that produces the 
 twist-2  PDF,  but it  also has the contaminating 
 structures containing $z^\alpha$,  $z^\beta$ or $g^{\alpha \beta}$.  When one takes,  as usual,
 $\alpha= \beta =+$  and $z= (z_-,z_+)$, the $z$-structures and $g^{\alpha \beta}$
 do not contribute to the matrix element of the operator ${\cal O}_g^{++}$ defining the gluon PDF.
 In case of the quasi-distribution, the  
 contaminating structures containing $z_3$ are  avoided when 
we  take
 $\alpha=0, \, \beta=0$
 (again, another definition of the gluon PQD corresponding to 
 $\alpha=3, \, \beta=3$ was chosen in Ref. \cite{Ji:2013dva} ).  
 Still, there remains contamination from the  $g^{\alpha \beta}$ structure and 
the momentum sum rule for gluons 
 \begin{align}
 \int_{-\infty}^\infty dy\,  y\, Q^{\rm soft}_g(y, P)  =
  \int_{-1}^1 dx\,  x \, 
f^{\rm soft}_g(x) + {\cal O} (\Lambda^2/P^2) 
 \
 \label{Qin8123gluon}
\end{align} 
is spoiled  by the $ {\cal O} (\Lambda^2/P^2)$ term
 brought in  by the $g^{\alpha \beta}$ admixture. 

\subsection{Primordial TMDs}

One may notice that  the ${\cal O}^\alpha (0,z;A)$ operator 
involves a straight-line link from $0$ to $z$ rather than a
stapled link usually used in the definitions of TMDs appearing  in the description of
Drell-Yan and semi-inclusive DIS processes.   As is well-known, the stapled links 
reflect initial or final state interactions inherent in  these processes.
The ``straight-link'' TMDs, in this sense, describe the structure
of a hadron when   it   is in 
 its  non-disturbed or ``primordial''  state. While it is unlikely that such a TMD can be measured
 in a scattering experiment, it is a well-defined  QFT object,
 and one may hope that it can be measured on the lattice
 through its connection (\ref{QTMD}) 
 to the quasi-distributions.

  \setcounter{equation}{0} 
\section{Models for soft part}

Let us now  discuss  some explicit  models  of the 
$k_\perp$ dependence of  soft TMDs    ${\cal F} (x, k_\perp^2)$.
In general, they are   functions
of two independent variables $x$ and $k_\perp^2$.
For simplicity, we will consider here the  case of     factorized models
  \begin{align}
{\cal F} (x, k_\perp^2) = f(x) \, \psi (k_\perp^2)  \  ,
\end{align} 
in which   $x$-dependence and  $k_\perp$-dependence appear in  separate factors.
Since, with our definitions, the  relations between VDFs and  TMDs  are the same in 
scalar and  spinor cases, we will refer  for brevity to scalar operators.

\subsection{Gaussian model}

It is  popular to assume  a Gaussian dependence on
$k_\perp$, 
  \begin{align}
{\cal F} _G (x, k_\perp^2) = \frac{f (x)}{\pi \Lambda^2}  e^{-k_\perp^2/\Lambda^2} \ . 
\label{Gaussian}
\end{align} 
Writing 
\begin{align}
{\cal F}_G (x, k_\perp^2)=&\frac{ {f(x)} }{2\pi^2 i \Lambda^2  }
\int_{-\infty}^{\infty} \frac{  \, d \sigma }{\sigma -i \Lambda^2} \, 
 \,  
 e^{- i k_\perp ^2 /\sigma} 
 \  ,  \label{gausssigma} 
\end{align} 
we see that the integral here involves both positive and
negative $\sigma$, i.e. formally  $ {\cal F} _G (x, k_\perp^2) $ 
cannot be written in the VDF representation 
(\ref{TMDsig}).  This is a consequence 
of the fact that 
the analytic continuation of $ {\cal F} _G (x, k_\perp^2) $
into the  region of   negative $k_\perp^2$
has an exponential  increase. 

However, since we are interested in positive $k_\perp^2$ only,  in our modeling 
we will just use the conversion formula (\ref{QTMD}) for 
all $k_\perp^2$ profiles for which it gives   convergent results.
For the Gaussian model we have then
 \begin{align}
 Q_G(y, P)  = &\frac{P}{\Lambda \sqrt{\pi} }  \,
 \int_{-1}^1 dx\,  %
f(x) \, 
  e^{- (x -y)^2 P^2 / \Lambda^2 }
 \  . 
 \label{QinG}
\end{align}

\subsection{Simple non-Gaussian models} 

 In the space of  impact parameters $z_\perp$, the   Gaussian 
 model gives a $e^{-z_\perp^2\Lambda^2/4}$ fall-off, and  one may  argue
 that the decrease 
is too fast   for large $z_\perp$.  In particular, 
 propagators $D^c (z,m)$ of massive particles have an exponential 
 $e^{-m|z|}$ fall-off  for spacelike intervals $z^2$.

To build models for TMDs  that resemble more  closely  the 
perturbative propagators in the deep spacelike region, we 
 recall that the propagator of a scalar particle 
with mass $m$ may be written as 
\begin{align} 
D^c(z,m) = &  \frac1{(4 \pi)^2}  \int_0^{\infty} e^{-i \sigma z^2/4 - i  (m^2 - i\epsilon)/\sigma  }  
 {d \sigma}  \  .
\label{alpharD}
\end{align}
It is the mass term that assures that the propagator falls off exponentially $\sim e^{-|z| m}$
for large  spacelike 
distances. 
At small intervals $z^2$, however,  the free particle propagator 
has a $1/z^2$ singularity while we want the soft part of 
$\langle p| \phi (0) \phi (z) |0 \rangle$ 
to be finite at $z=0$. The simplest way is to add a constant term $(-4/\Lambda^2)$ to $z^2$ 
in the VDF  representation  (\ref{newVDFx}). So, we take


\begin{align} 
\Phi (x, \sigma) = \frac{f(x)}{ 2i {m}{\Lambda} K_1 (2 m/ \Lambda) } e^{i \sigma /\Lambda^2  - i  m^2 /\sigma -\epsilon \sigma }  
\label{alpharDm}
\end{align}
as a model for the VDF, where  $K_1$  is the  modified Bessel function.
The sign of the $\Lambda^2$ term is fixed 
from the requirement that $(4/\Lambda^2-z^2)^{-1}$ should not 
have singularities for space-like $z^2$. 
This model 
corresponds to  the  following TMD 
\begin{align}
& {\cal F}_m(x, k_\perp^2 ) = f (x)\,  \frac{K_0 \left (2 \sqrt{ k_\perp^2 +m^2 } / \Lambda \right ) }{ \pi  m \Lambda
 K_1 (2 m/ \Lambda) }  \  . 
 \label{psim}
\end{align}
It  is finite for \mbox{$k_\perp =0$} reflecting  the  exponential $\sim e^{-m |z_\perp|}$
fall-off for  large $z_\perp$. 
To  avoid a two-parameter modeling,  one may take  $m=0$, i.e.
\begin{align} 
\Phi_{m=0}  (x, \sigma) = {f (x)} \,  \frac{e^{i \sigma /\Lambda^2  
  -\epsilon \sigma} }{ i \Lambda^2 } \ ,
\label{alpharD0}
\end{align}
which  corresponds to 
\begin{align}
& {\cal F}_{m=0}   (x, k_\perp^2) = 2 f(x)\,  \frac{K_0 ( 2 |k_\perp| / \Lambda) }{ \pi  \Lambda^2 } 
\  .
\end{align}
It has a logarithmic singularity for small $k_\perp$
that reflects a too slow $\sim 1/(1+z_\perp^2 \Lambda^2/4)$ fall-off  
for large  $z_\perp$.   
For the quasi-distribution,  we have 
 \begin{align}
 Q_{m=0}    (y, P)  = &\frac{P}{\Lambda}  \,
 \int_{-1}^1 dx\,  %
f(x) \, 
  e^{-2 |x -y| P / \Lambda } \  .
 \
 \label{Qinm}
\end{align} 

Note that   the Gaussian model and the $m=0$ models  have the same 
$\sim (1-z_\perp^2 \Lambda^2/4)$  behavior for small $z_\perp$, 
i.e. they correspond to the same value of the $\langle p | \varphi (0) \partial^2 \varphi (0)| p \rangle$
matrix element, provided that one takes the same value of $\Lambda$ in both models.
  For large $z_\perp$, however, the fall-off of the Gaussian model is 
   too fast, while that of the $m=0$ model is too slow. Thus, they  look like 
 two extreme  cases of one-parameter models, and 
we will  use them for illustration of the nonperturbative evolution 
of quasi-distributions, expecting that other models (e.g. $m \neq 0$  model)
will produce results somewhere in  between of  these two  cases.

 \setcounter{equation}{0} 
\section{Numerical results}

The full $-1\leq x \leq 1$ PDF-support segment 
 is usually split into the positive-$x$ ``quark'' region
and negative-$x$ ``antiquark''  region.  As we will see below,  the PQDs $Q(y,P)$ 
live on the whole $-\infty <y<\infty $ axis, even when they are 
generated from a TMD model that is non-zero for positive $x$ only.
Thus, to avoid confusion of what generates PQD for negative $y$,
 it makes sense to  separate the parts of PQDs coming  from 
 positive-$x$  and negative-$x$  parts of TMDs. 

To illustrate the pattern of the non-perturbative evolution
of quasi-distributions, we apply Eqs. (\ref{QinG}) and   (\ref{Qinm})  to a simple PDF 
$f(x)=(1-x)^3 \theta (0\leq x \leq 1)$    resembling 
nucleon valence distributions (an enthusiastic  reader  can easily 
obtain curves for more realistic  $(1-x)^3/\sqrt{x}$ valence models,
 for sea distribution models, etc.).

\begin{figure}[t]
    \centerline{\includegraphics[width=3.3in]{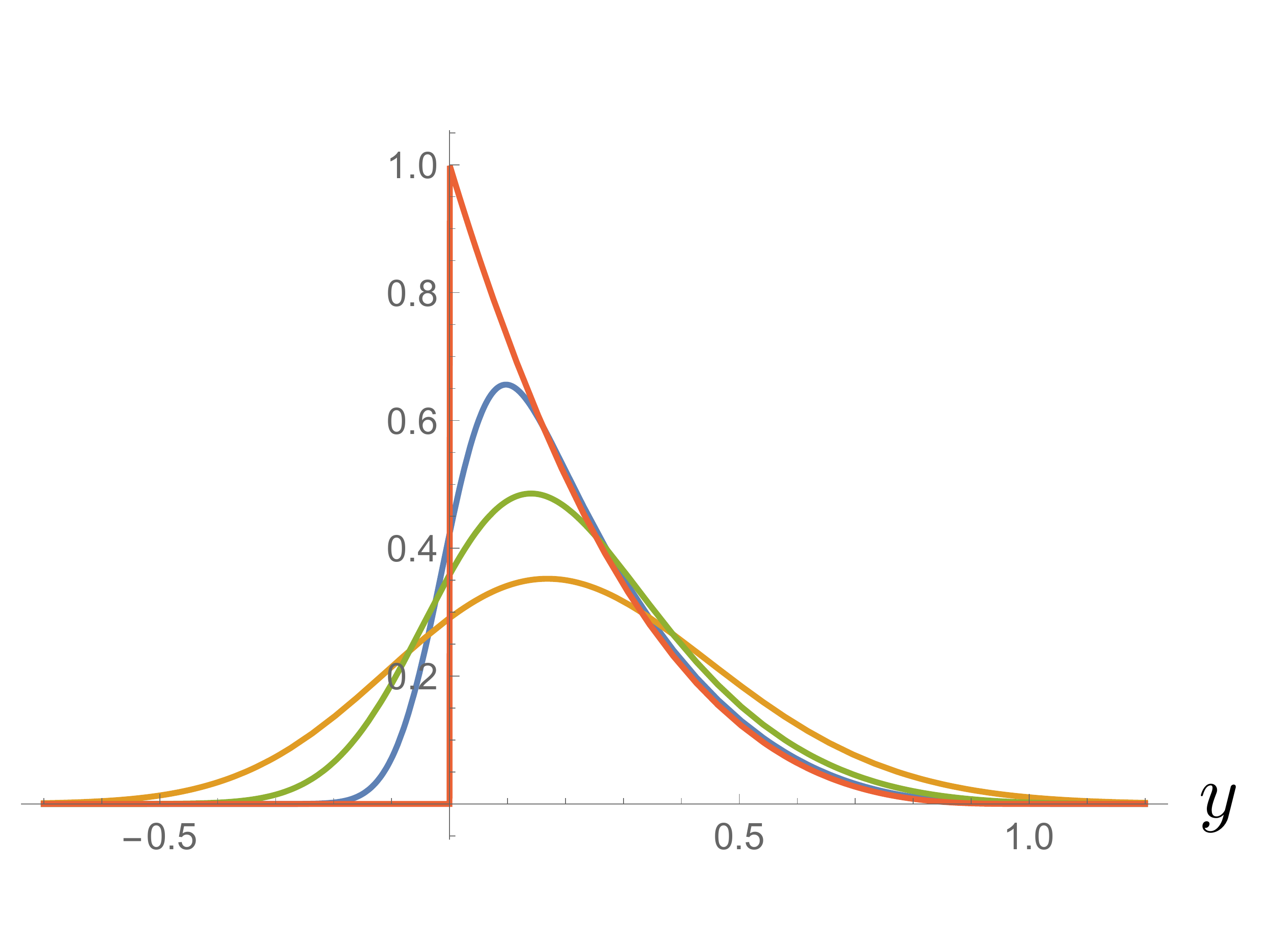}}
    \vspace{-0.5cm}
    \caption{Evolution of $Q(y,P)$  in the Gaussian  model  for {$P/\Lambda =3,5,10$} 
    (from bottom to top at $y=0.2$)
    compared to the  limiting PDF $f(y) = (1-y)^3 \theta (y)$.
    \label{Qgy}}
    \end{figure}

\begin{figure}[b]
    \centerline{\includegraphics[width=3.3in]{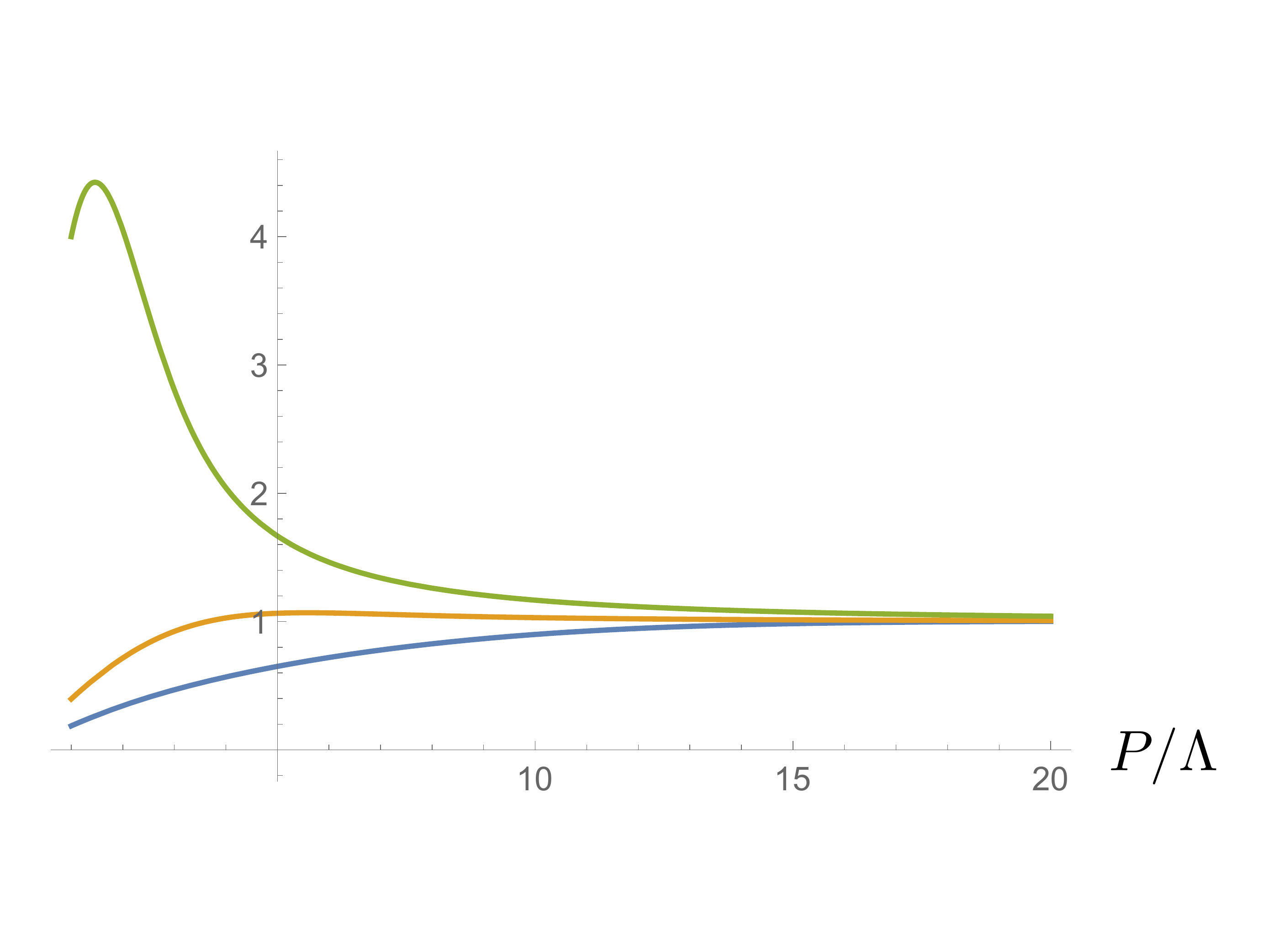}}
    \vspace{-0.5cm}
    \caption{Ratio  $Q(y,P)/f(y)$  in the Gaussian model  for $y =0.1,  0.3,  0.7$
    (from bottom to top)  and  \mbox{$f(y)=(1-y)^3$.}
    \label{QgP}}
    \end{figure}

\begin{figure}[t]
    \centerline{\includegraphics[width=3.3in]{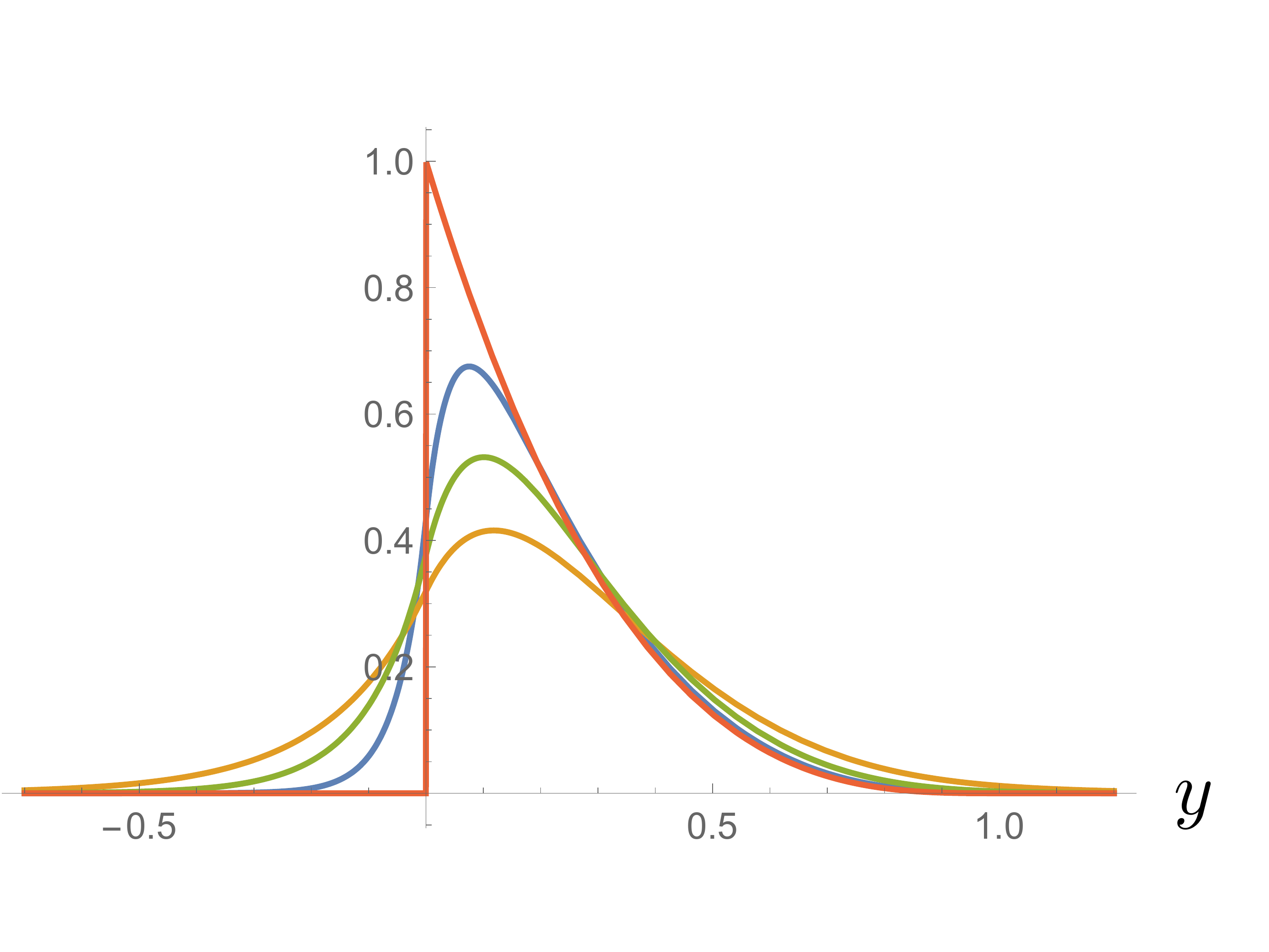}}
   \vspace{-0.5cm}
    \caption{Evolution of $Q(y,P)$  in the $m=0$ model  for {$P/\Lambda =3,5,10$}
   (from bottom to top at $y=0.2$)  compared to the  limiting PDF $f(y) = (1-y)^3 \theta (y)$.
    \label{Qfy}}
    \end{figure}

As one can see from Figs. \ref{Qgy}, \ref{Qfy}, the evolution patterns in our two models
 are very close to each other. 
 They  also resemble   the pattern observed in actual lattice calculations 
\cite{Lin:2014yra}-- \cite{Alexandrou:2016tjj}
 and in the diquark spectator model  \cite{Gamberg:2014zwa,Gamberg:2015opc,Bacchetta:2016zjm}.
 The  quasi-distributions are wider for small $P$, with their support visibly
 extending beyond the $0\leq y \leq 1$ segment, becoming narrower  (and higher in
 their maxima)  with increasing $P$.

\begin{figure}[b]
    \centerline{\includegraphics[width=3.3in]{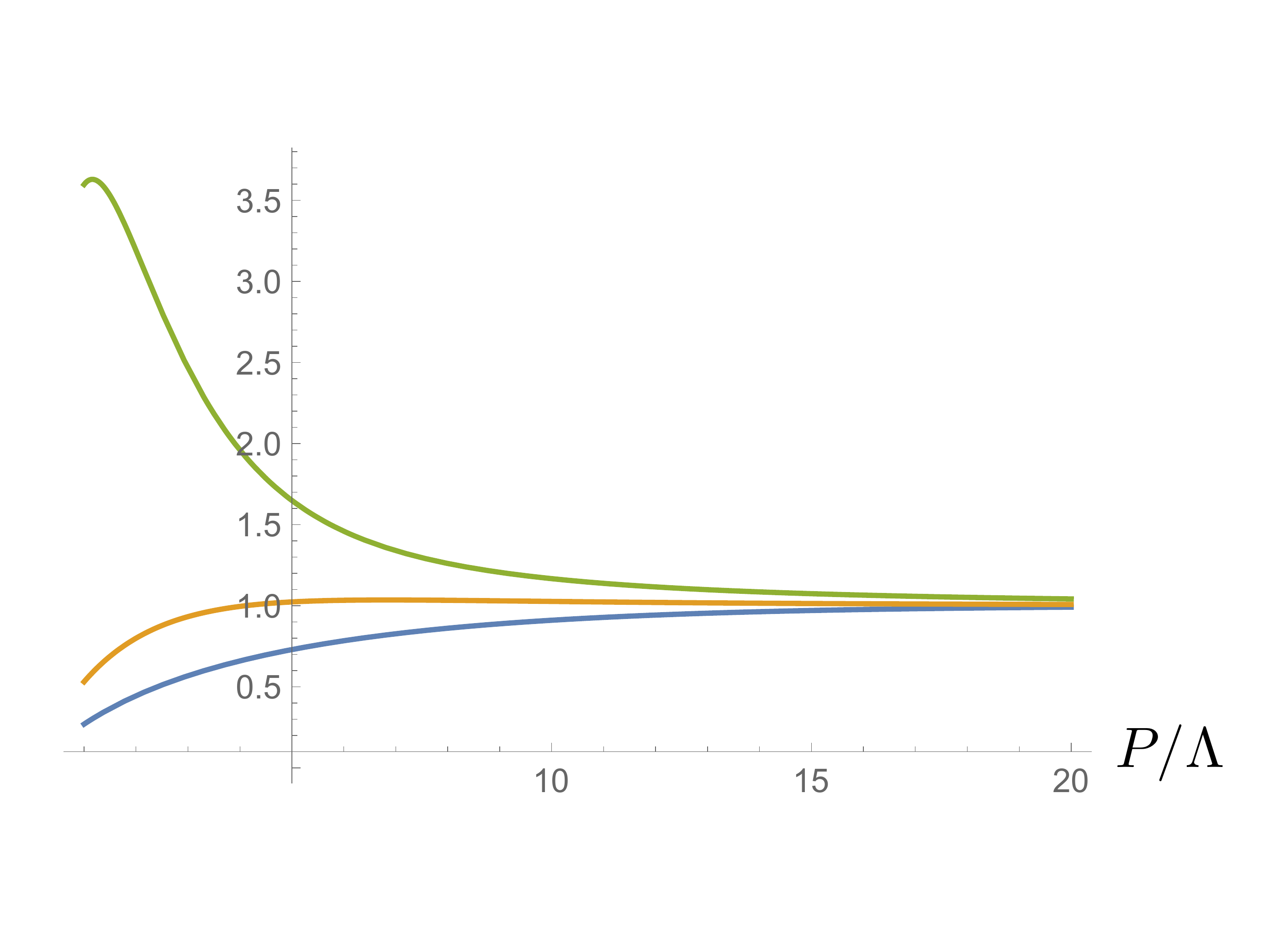}}
    \vspace{-0.7cm}
    \caption{Ratio  $Q(y,P)/f(y)$  in the $m=0$ model  for $y =0.1,  0.3,  0.7$ 
    (from bottom to top) and  \mbox{$f(y)=(1-y)^3$.}
    \label{QfP}}
    \end{figure}

The approach to the limiting $(1-y)^3$ shape is not uniform, as illustrated 
in Figs.  \ref{QgP}, \ref{QfP}.  For large $y=0.7$, the ratio $Q(y,P)/f(y)$ considerably 
exceeds 1  for small $P$ tending to the limiting value from above.
For   smaller $y=0.1$ and $y=0.3$, the ratio curves  tend to 1 from below.
One can see that $P/\Lambda \gtrsim 10$  is needed (or $P$ of the order of several GeV)
to get $Q(y,P)/f(y)$  close to 1 for these $y$ values.

 \setcounter{equation}{0} 

\section{Leading-order hard tail}

The nonperturbative evolution of $Q(y,P)$ essentially stops
for  $P/\Lambda \gtrsim 20$, and for larger values of $P$ the dominant role is played by  the perturbative evolution generated by the hard part.

The simplest  $\Phi  \sim 1/\sigma$  hard tail model  (\ref{Phihard})     
corresponds to a $\sim 1/k_\perp^2$ TMD. It is singular for $k_\perp=0$ while we want
TMDs be finite in this limit. The simplest regularization
\mbox{$ 1/k_\perp^2 \to 1/(k_\perp^2 +m^2) $}  corresponds to the change 
\mbox{$1/\sigma \to e^{-i m^2/\sigma}/\sigma$}  in the hard part of VDF, 
 \begin{align}
   \Phi^{\rm hard}  (x, \sigma) \, \to \frac{\Delta(x)}{\sigma}e^{- im^2/\sigma}  \,  
 \ . 
 \label{Phihardm}
\end{align}  
To proceed with the conversion formula, one needs the integral over $\sigma$
 \begin{align}
 I  (x,y, P)  = & \,\int_{0}^{\infty} \frac{d   \sigma}{ \sqrt{\pi \sigma}}
\frac{P}{\sigma}   \,   
  e^{- (x -y)^2 P^2 / \sigma- m^2/\sigma } \nonumber \\
  &=  \frac{1}{\sqrt{(x -y)^2+m^2/ P^2}}
 \ . 
 \label{hardsigma}
\end{align} 
This gives the hard part of a quasi-distribution
 \begin{align}
 Q^{\rm hard}  (y, P)  = &\,
 \int_{0}^1 dx\,  %
 \frac{\Delta (x)}{   
  \sqrt{(x -y)^2 +m^2/P^2} }
 \
 \label{deltaQ}
\end{align} 
($x>0$ is taken for definiteness) 
generating evolution with respect to $P^2$ in the form 
\begin{align}
P^2 \frac{d}{dP^2}  Q^{\rm hard}   (y, P)  = & \, \frac{ m^2}{2 P^2} 
 \int_{0}^1 dx\,  %
 \frac{\Delta (x)}{   
  [{(x -y)^2 +m^2/P^2}]^{3/2} }
 \ . 
 \label{evoQ}
\end{align} 
In  the $m/P \to 0$ limit we have 
\begin{align}
 \frac{ m^2}{2 P^2} 
 \int_{0}^1 dx\,  %
 \frac{P (x/z)}{   
  [{(x -y)^2 +m^2/P^2}]^{3/2} }= \,P (y/z)   +
 {\cal O} (m^2/P^2)
 \ , 
 \label{evoQ2}
\end{align} 
i.e. for large $P^2$ the quasi-distributions evolve according to the 
 perturbative evolution  equation with respect to  $P^2$.
 
The pattern of the sub-asymptotic $m^2/P^2$  dependence for the hard part may be illustrated 
by taking  $P(x/z)\to 1$. Then 
\begin{align}
   \, \frac{ m^2}{2P^2} &
 \int_{0}^1 dx\,  %
 \frac{1}{   
  [{(x -y)^2 +m^2/P^2}]^{3/2} } \nonumber \\ &= \frac12 
   \left(\frac{y}{\sqrt{ y^2+m^2/P^2
  }}+\frac{1-y}{\sqrt{   (1-y)^2 + m^2/P^2
}}\right)
   \nonumber \\ & =
 \theta (0 \leq y \leq 1) +{\cal O} (m^2/P^2) \   .
 \
 \label{evoQ2m}
\end{align} 

\section{Conclusions}

In this paper, we applied the formalism of parton virtuality distributions 
to study the $p_3$-dependence of quasi-distributions $Q(y,p_3)$.
We established a simple relation between PQDs and TMDs  that allows 
to derive models for PQDs from the models for  TMDs.
Our model results show a pronounced  nonperturbative 
evolution of PQDs for small and moderately large values of $p_3$
reflecting the transverse momentum dependence of TMDs,
i.e. the spatial structure of the hadrons.
Using two rather different models for the $k_\perp$ dependence 
of TMDs, we obtained very similar patterns of the $p_3$ 
dependence of PQDs $Q(y, p_3)$ for each particular $y$.
This observation may be used for a guided extrapolation 
of the moderate-$p_3$ lattice results to the $p_3 \to  \infty  $ limit. 
The  basic idea is to find  analytic models  for soft TMDs  that would 
successfully fit lattice PQDs for  several values of $p_3$,
and  then take the  $p_3 \to \infty$ limit. A practical implementation 
of this program should be  a subject of future studies.

Summarizing,  the study of  PQDs on the lattice in the domain 
of strong nonperturbative effects opens a new perspective in  investigations 
of  the three-dimensional structure of the hadrons.

\section*{Acknowledgements}

This work is supported by Jefferson Science Associates,
 LLC under  U.S. DOE Contract \#DE-AC05-06OR23177
 and by U.S. DOE Grant \#DE-FG02-97ER41028.


\end{document}